%% file: FastQuantumAlgo.tex
\documentclass{llncs}

\usepackage{a4wide}
\newcommand\F{\mathds{F}}
\newcommand\PS{{\tt PoSSo}$_q$}

\newcommand{\vectx}[2]{#1_1,\dotsc,#1_{#2}}
\usepackage{makeidx,dsfont,amsmath,algorithm,algpseudocode,comment}  
\newcommand\MQ{{\tt MQ}$_q$}
\newcommand\MQb{{\tt MQ}$_2$}

\newcommand{\polring}[3]{#1[\vectx{#2}{#3}]}

\newcommand{\Mac}[1]{\ensuremath{\mathcal{M}^{\mathrm{acaulay}}_{#1}}}
\newcommand{\FX}{\mathds{F}_2[x_1,\ldots,x_n]}
\newcommand{\dwit}{d_{{\rm wit}}}

\usepackage{hyperref}

\usepackage{tikz}
\usetikzlibrary{shadows}
\tikzstyle{fun}=[draw,very thick,fill=white,drop shadow]
\newcommand{\ket}[1]{\vert #1 \rangle}

\begin{document}
\frontmatter          
\pagestyle{plain}
\addtocmark{Quantum Computing} 

\mainmatter              
\title{Fast Quantum Algorithm for Solving Multivariate Quadratic Equations}
\titlerunning{Quantum MQ}  
%
\author{Jean-Charles Faug\`{e}re\inst{2,1} \and Kelsey Horan\inst{3} \and Delaram Kahrobaei\inst{3, 4} \and
Marc Kaplan\inst{1, }\inst5 \and Elham Kashefi\inst{1, }\inst{5} \and Ludovic Perret\inst{1,2}}

\authorrunning{Faug\`{e}re et al.} 
%
%
\institute{
 UPMC Univ Paris 06, UMR 7606, LIP6, F-75005, Paris, France\\
 CNRS, UMR 7606, LIP6, F-75005, Paris, France\\
\and
 INRIA, Paris Center, \\
\email{Jean-Charles.Faugere@inria.fr}, \email{ludovic.perret@lip6.fr}
\and PhD Program in Computer Science, The Graduate Center, The City University of New York \\ 365 5th Ave, New York, NY 10016, USA \\
\and New York University, Tandon School of Engineering, Brooklyn, NY 11201, USA\\
\email{khoran@gradcenter.cuny.edu}, \email{dkahrobaei@gc.cuny.edu}
\and School of Informatics \\ University of Edinburgh, 10 Crichton Street, Edinburgh EH8 9AB, UK \\ \email{kapmarc@gmail.com}, \email{ Elham.Kashefi@lip6.fr}}
\maketitle              

\begin{abstract}
In August 2015 the cryptographic world was shaken by a sudden and surprising announcement by the US National Security Agency  ({\tt NSA}) concerning plans to transition to post-quantum algorithms. Since this announcement post-quantum cryptography has become a topic of primary interest for several standardization bodies. The transition from the currently deployed public-key algorithms to post-quantum algorithms has been found to be challenging in many aspects. In particular the problem of evaluating the quantum-bit security of such post-quantum cryptosystems remains vastly open. Of course this question is of primarily concern in the process of standardizing the post-quantum cryptosystems.  In this paper we consider the quantum security of the problem of solving a system of  {\it $m$ Boolean multivariate quadratic equations in $n$ variables} (\MQb); a central problem in post-quantum cryptography. When $n=m$, under a natural algebraic assumption, we present a Las-Vegas quantum algorithm solving \MQb{} that requires the evaluation of, on average, $O(2^{0.462n})$ quantum gates. To our knowledge this is the fastest algorithm for solving \MQb{}.  
\keywords{Multivariate Quadratic Equations, Quantum Computation, Quantum Complexity}
\end{abstract}
\section{Introduction}
%
\input{intro.tex}

%
%
%

%
\section{Preliminaries}\label{prelim}
In the following we assume familiarity with standard classical and quantum computational notation, such as the standard bra-ket notation for specifying a quantum state. We use the following subsections to overview the classical and quantum algorithms which will be of use in this paper.
\subsection{Classical {\tt BooleanSolve}} \label{BooS}
\input{BooleanSolve.tex}

\subsection{Grover's Algorithm}
\label{sec:GroverAlgorithm}

\input{GroverAlg.tex}

\subsection{Quantum Gates}
\label{sec:QuantumGates}

\input{QuantumGates.tex}



\section{A Quantum Version of {\tt BooleanSolve}} \label{qbs}

\subsection{ {\tt QuantumBooleanSolve}}\label{sec:GroverBooleanSolve}
\input{qbs.tex}

\subsection{Quantum Oracle} \label{rs}
\input{classicalBooleanSolve.tex}


\input{QuantumOracle.tex}

\section{Complexity Analysis}
\label{sec:complexityanalysis}

\input{complexityanalysis.tex}

\section{Acknowledgment}
The first and last authors are partially supported by the french Programme d'Investissement d'Avenir under national project RISQ1 P141580\footnote{\url{https://risq.fr/?page_id=31&lang=en}}. 
Delaram Kahrobaei is partially supported by an  ONR (Office of Naval Research) grant N00014-15-1-2164, as well as a PSC-CUNY grant from the CUNY Research Foundation.
We also would like  to thanks the referees of PKC'18 for their comments on the first version of this document.

\bibliographystyle{splncs03}
\bibliography{Bib/FastQuantumAlgo.bib,Bib/bib_exter.bib,Bib/crypto_crossref,Bib/journal.bib,Bib/rankA.bib,Bib/rankB.bib,Bib/crypto_crossref}

\end{document}

%% file: intro.tex
The goal of this paper is to study the complexity of solving {\it systems of Boolean multivariate quadratic equations} (\MQb) in the quantum setting. This classical NP-hard  problem \cite{GJ79} is stated as follows:\smallskip 

\MQb{}\\
\textbf{Input.} $f_1(\vectx{x}{n}),\ldots,f_m(\vectx{x}{n}) \in \F_2[x_1,\ldots,x_n]$.\\
\textbf{Goal.} Find -- if any -- a vector $(\vectx{z}{n}) \in \F_2^n$ such that:
$$
f_1(\vectx{z}{n})=0,\ldots,f_m(\vectx{z}{n})=0.
$$ 
\MQb{} is a fundamental problem with many applications in cryptography, coding theory and beyond. Typically, the security of multivariate schemes is directly related to the hardness of \MQb{}, e.g. \cite{C:FauJou03,EC:KipPatGou99,PKC:BetFauPer11,DBLP:journals/jsc/BerbainGP09,DBLP:conf/asiacrypt/ChenHRSS16,cryptoeprint:2017:680}.  \MQb{} is then central to evaluating the security of such multivariate cryptosystems.  Besides multivariate cryptography, the security of a wide variety of cryptosystems is related  to  \MQb{}, via algebraic cryptanalysis \cite{DBLP:books/hal/Perret16}. This includes {\it post-quantum cryptosystems} \cite{opac-b1128738} such as code-based cryptography \cite{faugere:hal-01064687,FOPT10}, lattice-based cryptography  \cite{DBLP:conf/icalp/AroraG11,DBLP:journals/iacr/AlbrechtCFFP14}, $\ldots$  \smallskip      
  
The status of post-quantum cryptography is currently completely evolving. It is quickly moving from a purely academic theme to a topic of major industrial interest. This is mainly driven by the fact that post-quantum cryptography has recently received much attention from the standardization and policy sectors.  The triggering event appears to be the announcement in August $2015$ by the National Security Agency ({\tt NSA}) 
of  preliminary plans to transition the existing systems to quantum resistant algorithms\footnote{\url{https://www.nsa.gov/ia/programs/suiteb_cryptography/}}:
\begin{quote}``\emph{Currently, Suite B cryptographic algorithms are specified by the National Institute of Standards and Technology ({\tt NIST}) and are used by {\tt NSA}'s Information Assurance Directorate in solutions approved for protecting classified and unclassified National Security Systems ({\tt NSS}). Below, we announce preliminary plans for transitioning to {\bf quantum resistant}  algorithms.}''
\end{quote}
This was quickly followed by an announcement by \href{https://www.nist.gov/}{{\tt NIST}}, detailing the transition process \cite{NISTPQ}. {\tt NIST} then released in January $2016$ a call to select standards for post-quantum public-key cryptosystems: public-key exchange, signature and public-key encryption \cite{NISTsubmit}. The threat to see a large computer in a medium term was considered to be sufficient by {\tt NIST} to organize a renewal of the public-key cryptosystems deployed in practice.
\smallskip      
  
A key issue for the wide adoption of quantum-safe standards in the future is our confidence in their security. There is, therefore, a great need to develop {\it quantum cryptanalysis} against post-quantum cryptosystems. It is clear that a challenge in the next years will be to precisely evaluate the {\it quantum-bit security} of post-quantum cryptosystems submitted to the {\tt NIST} standardization process.\smallskip      

We study here how quantum techniques can be used to improve the complexity of solving \MQb{}; an important problem in post-quantum cryptography. In \cite{DBLP:journals/siamcomp/BennettBBV97}, the authors provide a theoretical upper limit  on the speed-up that can be obtained in the quantum setting. They demonstrated that  -- relative to an oracle chosen uniformly at random -- a problem in {\rm NP} can not be decided by any quantum algorithm in $o(2^{n/2})$. On the other hand, Grover's algorithm \cite{grover1996fast} is a quantum algorithm than can decide any problem of {\rm NP} in $O(2^{n/2})$; including  \MQb{}. Thus, Grover's algorithm is essentially optimal in the  setting of \cite{DBLP:journals/siamcomp/BennettBBV97}. We emphasize that this does not rule out the possibility of a greater than quadratic speed-up in the quantum setting. However, it is mandatory to take advantage of the problem structure to achieve this. \smallskip 

In this paper, we present an algorithm that beats the $O(2^{n/2})$ bound for solving \MQb{}. To do so, we combine Grover's technique with a Gr\"obner basis-based algorithm.       
  
\subsection{State of the Art}
\subsubsection{Classical Setting.} The question of solving \MQb{} has been investigated with various algorithmic techniques in the literature. We list below those techniques with the best asymptotic complexity.        
\paragraph{Exhaustive search.} The first, most obvious, technique for solving \PS{} is exhaustive search.
For $q=2$,  the authors of \cite{CHES:BCCCNS10} describe a fast exhaustive search for \MQb{} and provide the exact cost of this approach :  
$$
4 \,  \log_{2}(n) \,  2^n \mbox{ binary operations}. 
$$
A classical (and challenging) theme for \MQb{} is to design algorithms that are asymptotically faster than exhaustive search, i.e. that beat the $O(2^n)$ barrier.    
\paragraph{Approximation algorithm.}  Recently, the authors of \cite{NewMQ} proposed new techniques which solve \MQb{} faster than a direct exhaustive search.  The techniques from \cite{NewMQ} allows for the approximation of a system  $F=\big(f_1(\vectx{x}{n}),\ldots,
f_m(\vectx{x}{n})\big) \in \F_2[x_1,\ldots,x_n]$ by a single, high-degree, multivariate polynomial $P$ over $n'<n$ variables. The polynomial $P$ is constructed such that it vanishes on the same zeroes as the original system $F$ with high probability. We then must perform an exhaustive search on $P$ to recover, with high probability, the zeroes $F$. This leads to an algorithm for solving \MQb{} with complexity 
$$
O^*\big(2^{0.8765 \, n}\big).
$$ 
The notation $O^*$ omits polynomial factors.

\paragraph{Hybrid approaches.}  To date, the best methods for solving \MQb{} are based on Gr\"obner 
bases \cite{BCLA82,DBLP:journals/jsc/Buchberger06a}. More precisely, the fastest methods are hybrid techniques 
which combine exhaustive search and Gr\"obner bases algorithms \cite{FBP12,BFP09b,bardet2013complexity}. {\tt BooleanSolve}, an algorithm originally presented in \cite{bardet2013complexity}, falls into this category and is the asymptotically fastest approach to solving \MQb (Section \ref{BooS}). When $m=n$, the deterministic variant of  {\tt BooleanSolve} has complexity bounded by $O(2^{0.841n})$, while a Las-Vegas variant has expected complexity 
 $$
 O(2^{0.792  n}).
 $$ 
We emphasize that all stated complexities for {\tt BooleanSolve} are obtained under the assumption of a natural algebraic hypothesis on the input system. 
In contrast, the complexities of \cite{CHES:BCCCNS10,NewMQ} do not rely on any such assumption. 

\subsubsection{Quantum Setting.} 
The hardness of \MQ{} has been directly considered in \cite{westerbaan2016solving}, and somewhat indirectly in \cite{DBLP:conf/asiacrypt/ChenHRSS16}.

\paragraph{Quantum exhaustive search.} In \cite{westerbaan2016solving}, the authors proposed simple 
quantum algorithms for solving \MQb{}. The principle is to perform a fast exhaustive search by using Grover's algorithm. The authors derive precise resource estimates for their algorithms, demonstrating that we can solve $m -1$ binary quadratic equations in $n-1$ binary variables using $O(m+n)$ qubits and requiring the evaluation of $O\big(m n^2 2^{n/2} \big)$ quantum gates. The authors also describe a variant using $O\big(n + \log_2(m)\big) $ qubits but with twice as many quantum gates required, when compared to the first approach. In essence, this work constructs a quantum oracle to be used along with amplitude amplification performed by Grover's algorithm. The oracle is fairly simple and takes advantage of the structure of the \MQb{} problem, developing a straightforward way to evaluate a system of equations on a superposition of all possible boolean variable assignments. Then, Grover's algorithm is utilized to amplify those inputs which satisfy all provided equations.

\paragraph{Quantum hybrid  approach.} The main goal of \cite{DBLP:conf/asiacrypt/ChenHRSS16} is to construct a multivariate signature scheme based on random instances of \MQb{} and \MQ{} (for field bigger than $q>2$). However, in order to derive secure parameters, the authors considered a quantum variant of the hybrid approach from \cite{FBP12,BFP09b} using Grover's algorithm. They used this approach to explicitly compute the quantum-bit security of random instances of \MQ{} for given parameters. However, the authors of \cite{DBLP:conf/asiacrypt/ChenHRSS16} do not provide the asymptotic complexity of their approach. In this paper, we provide such an asymptotic analysis and build our quantum algorithm on top of {\tt BooleanSolve}. It should be mentioned that {\tt BooleanSolve} is inspired, but different, from \cite{FBP12,BFP09b}. So, the quantum algorithm presented here is different from the one sketched in \cite{DBLP:conf/asiacrypt/ChenHRSS16}.     



\subsection{Organization of the Paper and Main Results}
\paragraph{Overview of the results.}  The main result of this paper is the fastest known quantum algorithm algorithm for solving \MQb{} (Section \ref{sec:GroverBooleanSolve}). More precisely: 
 \begin{theorem}[summarized from Section \ref{sec:complexityanalysis}]
There is a quantum algorithm that solves \MQb{} and requires to 
\begin{itemize} 
\item evaluate $O(2^{0.47n})$ quantum gates for the deterministic variant, 
\item evaluate an expected number of $O(2^{0.462n})$ quantum gates for the probabilistic variant.   
\end{itemize} 
\end{theorem}
\paragraph{Overview of the results.} 
A natural step towards developing a quantum algorithm for the \MQb{} problem which outperforms quantum exhaustive search via Grover's algorithm \cite{westerbaan2016solving} would be the quantization of a classical algorithm for \MQb{} which outperforms classical exhaustive search. A first candidate for such quantization is the approximation algorithm \cite{NewMQ} described above. The quantization of such algorithm  for use in Grover's algorithm requires building a quantum circuit. Unfortunately, a basic approach to quantize the approximation algorithm mentioned does not seem to be possible, even for \MQb{}. 

Fortunately, we have been able to quantize {\tt BooleanSolve} using amplitude amplification techniques \cite{grover1996fast,MR1947332}. Under a natural algebraic assumption the new algorithm beats quantum exhaustive search, i.e. $O(2^{n/2})$. This is arguably a significant complexity result for a central problem in post-quantum cryptography, but more generally in computer science. The originality of our algorithm is to 
instantiate Grover's algorithm with a non-trivial oracle that implements the quantum circuit corresponding essentially to a simplified Gr\"obner basis computation (Section \ref{rs}).  We construct the quantum circuit required to implement the simplified Gr\"obner basis computation.                  

\paragraph{Cryptographic implications.}
The complexity analysis is especially important for selecting parameters in multivariate cryptography. It shows that in order to reach a quantum security level of $2^s$, one has to consider an instance of \MQb{} with at least  $s/0.462=2.16 \cdot s$ variables. In the table below, we provide the minimal number of variables $n$ (second column) required to reach a precise  security level (first column)      
The public-key in a multivariate cryptosystem is usually given by set of boolean equations. 
We report in the last column the minimum size required for a given security level.        
\begin{center}
\begin{tabular}{|c|c|c|c|}
\hline 
quantum sec. level  & $n$  & $O(n^3)$\\ 
\hline 
64 & 139 &  167.36 KB  \\ 
\hline 
80 &  173 &   326,4 KB \\ 
\hline 
128 & 277 & 1.33 MB \\ 
\hline 
256 & 555 &  10.65 MB \\ 
\hline
\end{tabular} 
\end{center}
Finally, we mention that in the signature scheme from \cite{DBLP:conf/asiacrypt/ChenHRSS16}, the authors proposed to use an instance of \MQb{} with $n=m=256$ variables to achieve a quantum security level of $128$ bits. According to our new result, the quantum security is slightly less, i.e. $118$ bits.    

\paragraph{Organisation.} After this introduction, the paper is organized as follows. In Section \ref{prelim}, we first review the two main components of our quantum 
algorithm :  {\tt BooleanSolve} (Section \ref{BooS}) and Grover's algorithm (Section \ref{sec:GroverAlgorithm}). We describe the new quantum algorithm,  
{\tt QuantumBooleanSolve}, in Section \ref{sec:GroverBooleanSolve}. We construct the quantum circuit for a simplified Gr\"obner basis computation, used as Grover's oracle, in Section \ref{rs}. Finally, we derive in Section \ref{sec:complexityanalysis} the complexity of our algorithm.

%% file: BooleanSolve.tex

As explained in the introduction, {\tt BooleanSolve}  \cite{bardet2013complexity}  is the fastest asymptotic algorithm for \MQb{}. From now on, we will refer to this algorithm as  {\tt ClassicalBooleanSolve}. We will indeed present a quantum version of this algorithm, {\tt QuantumlBooleanSolve}, in Section \ref{qbs}.\smallskip

Essentially, {\tt ClassicalBooleanSolve} first specializes a subset of the variables $x_1,\ldots,x_k$ and then checks the consistency of the specialized system using {\it Macaulay matrices} (Definition \ref{mac}). If the specialized system is found to be consistent, the original algebraic system is determined to have no solution. If the specialized system is inconsistent then the algorithm conducts an exhaustive search on the remaining $n-k$ variables and recovers the  solutions for the \MQb{} instance.\smallskip 

We cover the more relevant aspects of the theory behind the algorithm in an effort to keep this paper self contained, and refer the reader to additional preliminary and theoretical information which can be found in the original work \cite{bardet2013complexity}.
\begin{definition}\label{mac} 
Let $f \in \FX$, and $\phi(f)$ be the square-free part of $f$, i.e. the reduction of $f$ modulo $\langle x_i^2-x_i \rangle_{1 \leq i \leq n}$. The {\bf Boolean Macaulay matrix} of degree $d$ for 
a set of polynomials $F=(f_1,\ldots,f_m) \in \FX^m$, denoted by $\Mac{d}(F)$, has the following structure: the rows are the coefficients of polynomials $\{\phi(tf_i)\}$ where $1 \leq i \leq m$, $\text{deg}(tf_i) = d, t$ is a square-free monomial, and the columns are the square free monomials in the polynomial ring of degree at most $d$ ordered descendingly with respect to Degree Reverse Lexicographic ({\rm DRL}) ordering.
\end{definition}
We recall below some bounds on boolean Macaulay matrices that will be useful in the complexity analysis.  
\begin{proposition}{(\cite{bardet2013complexity})}
Let $F=(f_1,\ldots,f_m) \in \FX^m$. Denote by $r_{{\rm Mac}}$ (resp. $c_{{\rm Mac}}, s_{{\rm Mac}}$) the number of rows (resp. columns, number of nonzero entries) of the associated boolean Macaulay matrix $\Mac{d}(F)$. If $1 \leq d$ \textless $\frac{n}{2}$, then
$$
c_{{\rm Mac}} \textless \frac{1 - x}{1 - 2x}\binom{n}{d}, \quad  r_{{\rm Mac}} \textless m \frac{x^2}{(1-2x)(1-x)}\binom{n}{d},\quad s_{{\rm Mac}} \textless mn^2 \frac{x^2}{(1-2x)(1-x)} \binom{n}{d}
$$ 
where $x = \frac{d}{n}$.
\end{proposition}
{\tt ClassicalBooleanSolve} \cite{bardet2013complexity}  is based on a  
fundamental property of Macaulay matrices. 
Let $F=(f_1,\ldots,f_m) \in \FX^m$ and  ${\mathbf M}=\Mac{d}(F)$ be the corresponding boolean Macaulay matrix in degree $d$. It holds that if the linear system 
$$
u \cdotp {\mathbf M} = (0,0,\ldots,0,1)
$$ 
has a solution then $F$ does not have a solution in $\mathds{F}_2^n$. This reduces the problem of deciding the consistency of non-linear equations to the problem of solving a linear system. \smallskip 

We now need to determine which degree of the Macaulay matrix should be considered. This degree  is the so-called {\it witness degree} defined below:    
\begin{definition}{(\cite{bardet2013complexity})}
Let $F=(f_1,\ldots,f_m) \in \FX^m$ and $I \subset  \FX$ be the ideal defined by $F$. We set:
\begin{eqnarray*}
I_{\leq d} = \big \{p \in   \FX \,   \vert  \, p \in I, {\rm deg}(p) \leq d \big \},\\
J_{\leq d} = \big \{p \in   \FX\,  \vert \, \exists h_1,\ldots,h_{m+n}, \forall i \in \{1,\ldots,m+n\}, {\rm deg}(h_i) \leq d-2,\\
 p = \sum_{i=1}^m h_if_i + \sum_{j=1}^n h_{m+j} (x_j^2-x_j) \big\}.
\end{eqnarray*}
The {\bf witness degree} for $F$, denoted $\dwit(F)$, is the smallest integer $d_0$ such that 
$$
I_{\leq d_0} = J_{\leq d_0} \mbox{ and } \left \langle \{ {\rm LM}(f) \vert f \in I_{\leq d_0} \} \right \rangle = {\rm LM}(I),
$$ 
where ${\rm LM}(f)$ is the leading monomial of the polynomial $f$ with respect to {\rm DRL}  ordering.
\end{definition}
Alternatively,  the witness degree for $F$ can be defined  as the degree where any polynomial in a (minimal) Gr\"{o}bner basis of the system is obtained as a linear combination of the rows of the Macaulay matrix in this degree.   Therefore, given $F=(f_1,\ldots,f_m) \in \FX^m$, the witness degree provides an upper bound on the degree $d_0$ of $\Mac{d_0}(F)$ required to adequately determine the consistency of $F$. 

Under some  algebraic assumptions, the witness degree can be computed explicitly  
from the Hilbert series:
\begin{equation}\label{HS}
{\rm HS}(m,n,k)=\frac{(1+t)^{n-k}}{(1-t)(1+t^2)^m}.
\end{equation}
The witness degree, denoted by $\dwit(m,n,k)$, is given by the index of the first nonzero coefficient of \eqref{HS}.\smallskip   

Now that we have  reviewed all necessary background information, we can present the algorithm from \cite{bardet2013complexity} for solving  \MQb{}. \par \medskip \noindent
{\tt ClassicalBooleanSolve}

\par \smallskip \noindent
{\bf Input}: $f_1,\ldots,f_m \in \FX^m$ with $\text{deg}(f_i) = 2$ for all $i \in \{1,\ldots,m\}$.
\par \noindent
{\bf Output}: All boolean solutions to $f_1=\ldots=f_m=0$

\par \medskip \noindent
\begin{algorithmic}[1]
\Procedure{{\tt ClassicalBooleanSolve(m,n,k)}}{}
\State $S = \{\}$
\State $d_0 \leftarrow \dwit(m,n,k)$ 


\For{$(a_{n-k+1},\ldots,a_n) \in \mathds{F}_2^k$}
\For{$i=1 \ldots m$}
\State $\tilde{f_i}(x_1,\ldots,x_{n-k}) \leftarrow f_i(x_1,\ldots,x_{n-k},a_{n-k+1},\ldots,a_n) \in \mathds{F}_2[x_1,\ldots,x_{n-k}]$
\EndFor
\State ${\mathbf M} \leftarrow \Mac{d_0}(\tilde{f_1},\ldots,\tilde{f_m})$
\If{$u \cdotp {\mathbf M} = r = (0,\ldots,0,1)$ is inconsistent, determined by the {\tt SparseLinearSystemSolver}}
\State $T =$ solutions of the system $\tilde{f_1}=\ldots=\tilde{f_m}=0$ found by exhaustive search
\State $S \leftarrow S \cup T$
\EndIf
\EndFor
\State Return $S$
\EndProcedure
\end{algorithmic}
There are two variants of {\tt ClassicalBooleanSolve} : deterministic and Las-Vegas. The only difference is on the algorithm used in {\tt SparseLinearSystemSolver}, presented in Section ~\ref{rs}, which can be deterministic or probabilistic. The computational complexity of {\tt ClassicalBooleanSolve} is lower bounded by the complexity of the consistency check of the Macaulay matrices in degree $d_0$. Therefore, a complete complexity analysis will merely determine the time required to complete the consistency check in term of the input parameters. This yields:
\begin{theorem}{ (\cite{bardet2013complexity}) }\label{bs:cplx}
Let $\theta, 2 \leq \theta \leq 3$ be such that any two $n \times n$ matrices \cite{DBLP:books/daglib/0031325} can be multiplied in $O(n^{\theta})$ operations in the underlying field.
For any $\epsilon >0$, and $\alpha \geq 1$ and sufficiently large $m =\lceil \alpha n \rceil$, the complexity of all tests of consistency of Macaulay matrices in ${\tt ClassicalBooleanSolve(m,n,k)}$ is:
\begin{itemize}
\item $O(2^{(1- \gamma + \theta F_{\alpha}(\gamma) + \epsilon) n})$ in the deterministic variant,
\item $O(2^{(1- \gamma +2 F_{\alpha}(\gamma) + \epsilon) n})$ in the probabilistic variant,
\end{itemize}
where $\gamma = 1 - \frac{k}{n}$, $F_{\alpha}(\gamma)= - \gamma \log_2(D^D(1-D)^{(1-D)})$ with $D=M(\frac{\alpha}{\gamma})$, and  
$$
M(x) = -x + \frac{1}{2} + \frac{1}{2} \sqrt{2x^2 - 10x - 1 + 2(x+2) \sqrt{x(x+2)}}.
$$ 
\end{theorem}
This complexity is obtained by evaluating the cost of checking the consistency of $2^k=2^{(1- \gamma)n}$ Macaulay matrices.    

To derive the asymptotic complexity, we need to assume a certain algebraic condition on the systems considered
during the algorithm. 
\begin{definition}
Let $F=(f_1,\ldots,f_m)$ be quadratic polynomials in $\FX$ and $(1-\gamma)n \leq n$.
The system $F$ is called {\bf $\gamma$-strong semi-regular} if both the set of its solutions and the set 
$$
\left \{(a_{n-k+1},\ldots,a_n) \in \mathds{F}_2^k \, 
\vert  \,
\dwit\big(F(x_1,\ldots,x_{n-k},a_{n-k+1},\ldots,a_n)\big)  >\dwit(m,n,k) \right \}
$$
have cardinality at most $2^{(1- \gamma +2 F_{\alpha}(\gamma) + \epsilon)}$, with $\epsilon >0$ and $F_{\alpha}$ as in  Theorem \ref{bs:cplx}.
\end{definition}
Under this assumption, we can now minimize, in term of $k$, the complexities of Theorem \ref{bs:cplx}.
The  results are provided for various values of $\theta$: $3$ which is the upper bound, $2.376$  which is the current best  theoretical bound \cite{DBLP:conf/issac/Gall14a}, and $2$ which requires careful consideration of the linear algebra problem.            
 \begin{lemma} 
Let the notations be as in Theorem \ref{bs:cplx}. The function 
$1-\gamma+\theta F_{\alpha}(\gamma)$ is bounded by: 
\begin{itemize}
\item $1-0.112 \alpha$, when $\theta=3$ and $\gamma=0.27 \alpha$,
\item $1-0.159 \alpha$, when $\theta=2.376$ and $\gamma=0.40 \alpha$,
\item $1-0.208 \alpha$, when $\theta=2$ and $\gamma=0.55 \alpha$.
\end{itemize}    
\end{lemma} 
Finally:   
\begin{theorem} 
{\tt ClassicalBooleanSolve} is correct and solves \MQb{}. If $m=n$, then the algorithm has complexity $O(2^{0.841n})$, if the system is $0.40$-strong semi-regular, for the deterministic variant, and of expectation $O(2^{0.792n})$, if the the system is $0.55$-strong semi-regular, for the Las-Vegas probabilistic variant.
\end{theorem}
This is essentially the cost of the first step, i.e. testing the Macaulay matrices, since the second step, i.e. exhaustive search, has negligible cost when compared to the consistency check.

%% file: GroverAlg.tex
Grover's algorithm \cite{grover1996fast}, often called \textit{database search}, is a quantum algorithm that can be implemented to reduce computation time for the exhaustive search of a function over the entire function domain. The problem solved by Grover's algorithm is as follows: given a function $f: \{0,1\}^n \rightarrow \mathds{F}_2$, determine the unique $x^* \in  \{0,1\}^n$ such that $f(x^*)=1$.

Determining such a $x^*$ with a classical computer requires exhaustive search on the entire function domain of $f$. Classical computation techniques cannot do better than evaluating $f$ over every possible input, resulting in time complexity of $O^*(2^n)$. Grover's quantum algorithm can determine $x^*$ with merely $2^{\frac{n}{2}}$ evaluations of $\mathcal{F}$, the quantum circuit which evaluates the function $f$. 

Grover's algorithm can be extended to perform exhaustive search over a function where $\vert f^{-1}(1) \vert = M$ with $M \geq 1$, as well as searching over a function where the preimage of $1$ has arbitrary size. Here we present a simple version of the algorithm.

In the quantum oracle model, when presented with a quantum oracle for the evaluation of $f$, the problem is to locate an $x^*$ such that $f(x^*)=1$. The algorithm utilizes two unitary operations. First, a rotation $O_f^{\pm}: \alpha_x \vert x \rangle \rightarrow (-1)^{f(x)} \alpha_x \vert x \rangle$ which flips the sign of the phase of the desired $x^*$. $$O_f^{\pm}:\frac{1}{\sqrt{2^n}}\sum_{x \in \{0,1\}^n} x \rightarrow \frac{1}{\sqrt{2^n}}\sum_{x \in \{0,1\}^n, x \neq x^*} \vert x \rangle - \frac{1}{\sqrt{2^n}} \vert x^* \rangle$$
Second, a diffusion operator $D$ which rotates the state around the average amplitude, $\mu=\frac{1}{2^n}\sum_{x \in 2^n} a_x$ of $x \in 2^n$, $$D:\sum_{x \in \{0,1\}^n} \alpha_x \vert x \rangle \rightarrow \sum_{x \in \{0,1\}^n} (2 \mu - \alpha_x) \vert x \rangle$$
Successive application of these two oracles performs amplitude amplification on the quantum computer, essentially taking the state of the computer from a uniform superposition over all inputs to a state that, when measured, with high probability will return $x^*$. To converge to such a final state the oracles $O_f^{\pm}$ and $D$ must be applied $\lceil \frac{\pi}{4} \sqrt{\frac{2^n}{\vert f^{-1}(1) \vert}} \rceil$ times.

The algorithm proceeds as follows: begin by using a Hadamard gate, $H^{\oplus n}$, to prepare the quantum computer in a uniform superposition over all possible inputs, $\frac{1}{\sqrt{2^n}}\sum_{x \in \{0,1\}^n} \vert x \rangle$. Following this, apply $O_f^{\pm}D$ to the quantum state $\lceil \frac{\pi}{4} \sqrt{\frac{2^n}{\vert f^{-1}(1) \vert}} \rceil$ times. Finally, measure to obtain $x^*$ with high probability. The computational complexity of Grover's algorithm is $O(2^n \cdotp \mathcal{F})$ where $\mathcal{F}$ is the complexity of the quantum oracle for $f$.

\begin{theorem}[Amplitude Amplification (\cite{MR1947332})]
Let $\mathcal{A}$ be a quantum algorithm that, with no measurement, produces a superposition $\sum_{x \in G} a_x \vert x \rangle + \sum_{y \in B} a_y \vert y \rangle$. Let $a = \sum_{x \in G} \vert a_x \vert^2$ be the probability of obtaining, after measurement, a state in the good subspace $G$. Then, there exists a quantum algorithm that calls $\mathcal{A}$ and $\mathcal{A}^{-1}$ as subroutines $O(\frac{1}{\sqrt{a}})$ times and produces an outcome $x \in G$ with a probability at least $\text{max}(a, 1-a)$.
\end{theorem}

The key to successfully performing Grover's algorithm for the function $f$ is to determine the quantum circuit for the function, in order to construct $O_f^{\pm}$. It is sufficient to provide an oracle that computes the function $f$, i.e. provide a unitary operator $U_f$ in the form of a quantum circuit which calculates $\vert x \rangle \vert y \rangle \rightarrow \vert x \rangle \vert y \oplus f(x) \rangle$, evaluating the function at a superposition of all possible inputs. Then, Grover's algorithm can be used to amplify the desired output for measurement. What remains is to show that the quantum analog of {\tt ClassicalBooleanSolve} is reversible and computable on a quantum computer. In the following section, we will construct the quantum circuit for the algorithm {\tt ClassicalBooleanSolve} and analyze the complexity of the circuit.

%% file: QuantumGates.tex
The following gates are quantum gates of interest which operate on qubits, each directly corresponding to reversible classical gates. For qubits $\vert x \rangle, \vert y \rangle, \vert z \rangle$ the gates perform the following operations:
\begin{itemize}
\item {\it CNOT} (XOR, Feynman) $$\text{CNOT}\vert x \rangle \vert y \rangle = \vert x \rangle \vert x+y \rangle$$

\item {\it Toffoli} (AND)
$$\text{T}\vert x \rangle \vert y \rangle \vert z \rangle = \vert x \rangle \vert y \rangle \vert z + xy \rangle$$
\item {\it X} (NOT)
$$\text{X} \vert x \rangle = \vert \bar{x} \rangle = \vert 1 + x \rangle$$
\item {\it n-qubit Toffoli} (AND)
$$\text{T}_n \vert x_1 \rangle \dotsc \vert x_n \rangle = \vert x_1 \rangle \dotsc \vert x_{n-1} \rangle \vert x_n + (x_1 \dotsc x_{n-1}) \rangle$$
\item {\it Swap} $$\text{S}\vert x \rangle \vert y \rangle = \vert y \rangle \vert x \rangle$$
\end{itemize}

It is important to note that $T_1 = X$, $T_2 =$ CNOT and $T_3 = T$. In terms of computational complexity, X, SWAP and CNOT gates are relatively cheap to compute, while the n-qubit Toffoli gates are more expensive; $T_n$ is equivalent to computing $2n$ CNOT gates. Accounting for these equivalences can change the reported computational complexity of a given circuit. Additionally, it is important to note that one can emulate a Toffoli gate over $\mathds{F}_q$ using $\log(q)$ basic $T$ gates, in order to determine the additional resources required for any general extension of quantum computations over $\mathds{F}_2$ to quantum computations over $\mathds{F}_q$.

%% file: qbs.tex
We explain here how to combine the {\tt ClassicalBooleanSolve} algorithm (Section \ref{BooS}) with Grover's algorithm  (Section \ref{sec:GroverAlgorithm}). {\tt ClassicalBooleanSolve} conducts two exhaustive searches over the variables. The first exhaustive search is over the last $k$ variables, specializing $x_{n-k+1},\ldots,x_{n}$ and projecting to the $k$ last components of a solution. The second exhaustive search, when necessary, is on the first $n-k$ variables, and allows the algorithm to determine the entire solution. It is clear that one can utilize Grover's algorithm, as in \cite{DBLP:conf/space/2016}, to quantize the second exhaustive search and obtain a speed up over the classical complexity. In what follows, ${\tt QuantumSearch}$ will refer to the quantum algorithm \cite{DBLP:conf/space/2016} that solves \MQb{}. We will see that Grover's algorithm can be used to speed-up the first exhaustive search as well. Essentially, we will quantize the consistency check on Macaulay matrices by providing a quantum circuit which can be used as the function oracle in Grover's algorithm.\smallskip  
 
Let $F=(\vectx{f}{m})\in\polring{\F_2}{x}{n}^m$. We consider the function 
 $F^{{\rm cons}}_{F,k} : \F_2^k \mapsto \{ 0,1 \}$ which evaluates $F$ on $(x_1,\ldots,x_{n-k},y_1,\ldots,y_k)$ with $(y_1,\ldots,y_k) \in \F_2^k$ and returns $1$ only if the non-linear system defined below is consistent :
\begin{eqnarray*} 
\tilde{F}&=&(\vectx{\tilde{f}}{m})\\
			&=&\big(f_1(x_1,\ldots,x_{n-k},y_1,\ldots,y_k)
,\ldots,f_m(x_1,\ldots,x_{n-k},y_1,\ldots,y_k)\big) \in \F_2^m[x_1,\ldots,x_{n-k}]
\end{eqnarray*}
This is reduced to check whether the linear system below has a solution: 
\begin{equation} \label{eqfonda}
u \cdotp \Mac{d}(\tilde{F}) = (0,0,\ldots,0,1), \mbox{ for a well chosen $d$.}
\end{equation} 
In order to quantize {\tt ClassicalBooleanSolve}, we then proceed in two steps. We first use Grover's algorithm, along with the quantum circuit which evaluates $F^{{\rm cons}}_{F,k}$ to determine ${\bf a}_2:=(a_{n-k+1},\ldots,a_n) \in \F_2^k$ such that $F^{{\rm cons}}_{F,k}({\bf a}_2)=1$, i.e. ${\bf a}_2$ is such that the non-linear system below is consistent: 
 $$
 \tilde{F}:=(\vectx{\tilde{f}}{m})=\big(f_1(x_1,\ldots,x_{n-k},{\bf a}_2),\ldots,f_m(x_1,\ldots,x_{n-k+1},{\bf a}_2)\big).
 $$
 The ${\bf a}_2$ then corresponds to the variable assignments of the last components in a solution for the system $F$. 
 We can then use Grover's algorithm, via {\tt QuantumSearch} on $\tilde{F}$ to find the remainder of a complete solution corresponding to ${\bf a}_2$. 
  
\par \medskip \noindent
\begin{algorithmic}[1]
\Procedure{{\tt QuantumBooleanSolve(m,n,k)}}{}
\State  ${\bf a}_2:=(a_{n-k+1},\ldots,a_n):={\tt GroverSearch}(F^{{\rm cons}}_{F,k})$
 \State $\tilde{F} \leftarrow (\vectx{\tilde{f}}{m})=\big(f_1(x_1,\ldots,x_{n-k},{\bf a}_2),\ldots,f_m(x_1,\ldots,x_{n-k+1},{\bf a}_2)\big)$
 \State ${\bf a}_1:=(a_1,\ldots,a_{n-k}):={\tt QuantumSearch}(\tilde{F})$
 \State Return $({\bf a}_1,{\bf a}_2)$
\EndProcedure
\end{algorithmic}
\medskip


The most essential part of determining the advantage of using Grover's algorithm to improve the computational complexity of {\tt ClassicalBooleanSolve} is to construct the quantum circuit for $F^{{\rm cons}}_{F,k})$. Below, we construct the quantum circuit that solves \eqref{eqfonda}.

%% file: classicalBooleanSolve.tex

{\tt QuantumBooleanSolve} consists of constructing a quantum oracle for the consistency check of Macaulay matrices, i.e. for $F^{{\rm cons}}_{F,k}$.  We provide the classical complete sparse linear system solver for classical consistency checks below, as presented by Giesbrecht et al. \cite{bananas}, and then provide an outline of the quantum circuit. Classical algorithm complexity is provided in a black-box model, where we assume access to a black-box for computing matrix-matrix and matrix-vector products. Then, the complexity is given as the number of calls to such black boxes, as well as the number of additional field operations required.
\subsubsection{Classical case.}  
{\tt SparseLinearSystemSolver} is the classical algorithm employed to determine the consistency of the Macaulay matrices in degree $d_0$. The algorithm takes as input a matrix $A$, a vector $b$ and a subset of the field or a field extension $\mathcal{L}$, and outputs either a solution $x$ to $Ax=b$ or a certificate of inconsistency, $u$. This classical algorithm requires $O(n)$ evaluations of black box algorithms for matrix-vector multiplications, as well as an additional $O(n^2 \log n \log \log n)$ additional field operations. Subroutines of the {\tt SparseLinearSystemSolver} can be found below.

\par \medskip \noindent
{\tt SparseLinearSystemSolver} \par \smallskip \noindent {\bf Input}: $A \in \mathds{F}^{n \times n}, b \in \mathds{F}, \mathcal{L} \subset \mathds{F}$ with $\vert \mathcal{L} \vert$ \textgreater $2n(n-1)$
\par \noindent
{\bf Output}: Any of the following $3$ return values, as an evaluation of the matrix $A$: {\it (nonsingular, $x$)} where $x = A^{-1}b$, {\it (singular-consistent, $x$)} with $x$ a random element of the solution space, {\it (singular-inconsistent, $u$)} with $u^tA=0$ and $u^t b \neq 0$, certifying the inconsistency of the system.

\par \smallskip \noindent
\begin{algorithmic}[1]
\Procedure{{\tt SparseLinearSystemSolver}}{}
\State $(\hat{f}(z),x) \leftarrow {\tt Wiedemann}(A,b,\mathcal{L})$
\If{$x \in \mathds{F}^{n \times 1}$ and $Ax=b$}
\State Return {\it (nonsingular,$x$)}
\EndIf
\State $\alpha_2,\alpha_3,\ldots,\alpha_n,\beta_2,\ldots,\beta_n,\gamma_1,\ldots,\gamma_n \xleftarrow{\$} \mathcal{L}$
\State \[ U = \begin{bmatrix}
    1       & \alpha_2 & \alpha_3 & \alpha_4 & \dots & \alpha_{n-1} & \alpha_n \\
    0      & 1 & \alpha_2 & \alpha_3 & \dots & \alpha_{n-2} & \alpha_{n-1} \\
        0      & 0 & 1 & \alpha_2 & \dots & \alpha_{n-3} & \alpha_{n-2} \\
    \hdotsfor{7} \\
    0      & 0 & 0 & 0 & \dots & 0 & 1
\end{bmatrix} \]
\State \[ L = \begin{bmatrix}
    1       & 0 & 0 & 0 & \dots & 0 & 0 \\
    \beta_2      & 1 & 0 & 0 & \dots & 0 & 0 \\
        \beta_3      & \beta_2 & 1 & 0 & \dots & 0 & 0 \\
    \hdotsfor{7} \\
    \beta_n      & \beta_{n-1} & \beta_{n-2} & \beta_{n-3} & \dots & \beta_{2} & 1
\end{bmatrix} \]
\State $B \leftarrow UAL$
\State $\hat{f}(z) \leftarrow {\tt Wiedemann}(B,0,\mathcal{L})$ the minpoly of $B$
\State $f(z) \leftarrow \frac{\hat{f}(z)}{z}$
\If{$z \vert f(z)$}
\State Repeat all steps following the random generation of $U,L$
\EndIf
\State $r \leftarrow \text{deg}(f)$
\State $c \leftarrow Ub$
\If{{\it (True,$u$)} $\leftarrow {\tt RandomSol}(B^t,0,f,\mathcal{L})$ and $u^tc \neq 0$}
\State Return {\it (singular-inconsistent, $u$)}
\EndIf
\If{{\it (True, $x$)} $\leftarrow {\tt RandomSol}(B,c,f,\mathcal{L})$}
\State Return {\it (singular-consistent, $x$)}
\EndIf
\State Return to Step 6
\EndProcedure
\end{algorithmic}

\par \medskip \noindent
{\tt RandomSol}: a subroutine of {\tt SparseLinearSystemSolver}, intended to return a random element of the solution space to the system $Ax=b$. The algorithm is stated to require $O(r)$ evaluations of the black-box for matrix-vector product, and $O(nr)$ additional field operations. \par \medskip \noindent
{\tt RandomSol}
\par \smallskip \noindent
{\bf Input}: $A \in \mathds{F}^{n \times n}, b \in \mathds{F}^{n \times 1}, f(z) \in \mathds{F}[z]$ with $f(0) \neq 0$, and $\mathcal{L} \subset \mathds{F}$ with $\vert \mathcal{L} \vert$ \textgreater $2n(n-1)$
\par \noindent
{\bf Output}: One of the following two return values: {\it (False)} indicating no solution, {\it (True, $\hat{x}$)} with $\hat{x}$ a random solution to the system $Ax=b$.

\par \smallskip \noindent
\begin{algorithmic}[1]
\Procedure{{\tt RandomSol}}{}
\State $w \leftarrow (w_1, \ldots, w_n)^t$ with $w_i \in_{\$} \mathcal{L}$
\State $r \leftarrow \text{deg}(f(z))$
\State $b' = (b'_1,\ldots,b'_r) \leftarrow b + Aw$, the first $r$ entries of the calculated vector $b'$
\State $A_r$, the leading $r \times r$ submatrix of $A$
\State $x \leftarrow -\sum_{i=1}^{n} \frac{f_i}{f_0}A_r^ib'$
\If{$Ax=b$}
\State Return {\it (True, $x$)}
\EndIf
\State Return {\it (False)}
\EndProcedure
\end{algorithmic}

\par \medskip \noindent
{\tt Wiedemann}: a subroutine of {\tt SparseLinearSystemSolver}. The deterministic version of the {\tt Wiedemann} algorithm is presented below, as seen in \cite{wiedemann1986solving}.
\par \medskip \noindent
{\tt Wiedemann}
\par \smallskip \noindent
{\bf Input}: $A \in \mathds{F}^{n \times n}$, $b \in \mathds{F}^{n \times 1}$, $\mathcal{L} \subset \mathds{F}$
\par \noindent
{\bf Output}: One of the following two return values: $x$ such that $x \in \mathds{F}^{n \times 1}$ and $Ax=b$ or $\hat{f}(z)$, a factor of $\text{minpoly}(A) \in \mathds{F}[z]$.

\par \smallskip \noindent
\begin{algorithmic}[1]
\Procedure{{\tt Wiedemann}}{}
\State $S = \{\}$
\For{i = 0 \ldots 2n-1}
    \State $S = S \cup \{A^ib\}$
\EndFor
\State $k \leftarrow 0$
\State $g_0(z) \leftarrow 1$
\While{$\text{deg}(g_k)$ \textless $n$ and $k$ \textless $n$}
    \State $u_{k+1} \leftarrow e_{k+1}$
    \State $s_k \leftarrow \{(u_{k+1},A^ib)\}_{i=0}^{2n-1}$
    \State $gs_k \leftarrow \{(u_{k+1},A^ig_k(A)b)\}_{i=0}^{2n-1-\text{deg}(g_k)}$ 
    \State $f_{k+1} \leftarrow \text{minpoly}(gs_k)$ using the Berlekamp-Massey algorithm
    \State $g_{k+1} \leftarrow f_{k+1}g_k$
    \State $k \leftarrow k + 1$
\EndWhile
\State Return $x = - \sum_{i=1}^{\text{deg}(g_k)}g_k[i]A^{i-1}b$
\EndProcedure
\end{algorithmic}

\par \medskip \noindent
From the presentation of the classical algorithm  {\tt SparseLinearSystemSolver}, and the subroutines, it is clear that the significant operations performed classically are matrix multiplication and vector-matrix products. The complexity analysis for such operations is given in a black-box model. Therefore, it is sufficient to show that these operations can be executed (in comparable time) by a quantum circuit on a quantum computer, in order to construct a quantum oracle for matrix consistency checking.

%% file: QuantumOracle.tex
\subsubsection{Quantum case.} 
As the basic operations implemented by the classical consistency check are linear algebra on matrices, it is important to verify that this linear algebra can be computed on a superposition of inputs via a quantum circuit. We must check that a reversible unitary operation can compute the required algebraic computations on a quantum computer, with comparable computational complexity to their classical analogs.

\paragraph{Equality Testing.}
Binary equality testing, i.e. checking if $b = \tilde{b}$, can be computed via one CNOT and one X gate, as follows: $\ket {b} \ket {\tilde{b}}  \ket {0} \rightarrow \ket {b} \ket {\tilde{b}} \ket {b \oplus \tilde{b} \oplus 1}$.

\paragraph{Matrix Vector Multiplication.}
The formula for matrix vector multiplication $Ax$ calculates each element of the product vector $b=(b_1,\ldots,b_n)=(a_{11}x_1+a_{12}x_2+a_{1n}x_n,a_{21}x_1+a_{22}x_2+\ldots+a_{2n}x_n,\ldots,a_{n1}x_1+a_{n2}x_2+\ldots+a_{nn}x_n)$. To compute each $b_i$ we require at most $n$ products between the elements of $A$ and of $x$. Therefore, the matrix-vector product requires at most $n^2$ products. It remains to show that the computation is reversible.

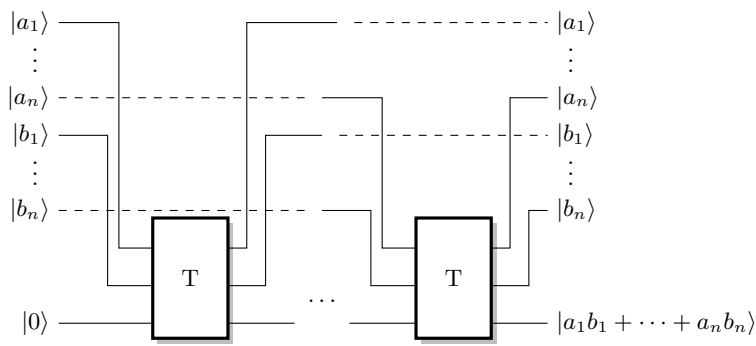
\begin{figure}[h]
\begin{center}
\begin{tikzpicture}
	\node[left]       (a1)    at (0.5,4) {$\ket {a_1}$};
	\node[left]       	    at (0.35,3.6) {$\vdots$};
	\node[left]       (anin)at (0.5,3) {$\ket {a_n}$};
	\node[left]       (an) at (4,3) {};
	\node[left]       (b1) at (0.5,2.5) {$\ket {b_1}$};
	\node[left]       	    at (0.35,2.1) {$\vdots$};
	\node[left]       (bnin) at (0.5,1.5) {$\ket {b_n}$};
	\node[left]       (bn) at (4,1.5) {};	
	\node[left]       (0input) at (0.5,0) {$\ket 0$};
	\node		 at (2,0) {} edge[-] (0input);
	\node[fun, minimum width=1cm, minimum height=1.6cm] at (2.25,0.6) {T};
	\node[fun, minimum width=1cm, minimum height=1.6cm] at (5.75,0.6) {T};
	\node[right]       (a1out)    at (4.2,4) {};
	\node[right]       (a1final)    at (7,4) {$\ket {a_1}$};
	\node[right]       	    at (7.15,3.6) {$\vdots$};
	\node[right]       (anout) at (7,3) {$\ket {a_n}$};
	\node[right]       (b1out) at (4,2.5) {};
	\node[right]       (b1final) at (7,2.5) {$\ket {b_1}$};
	
	\node[right]       	    at (7.15,2.1) {$\vdots$};
	\node[right]       (bnout) at (7,1.5) {$\ket {b_n}$};
	\node[right]       (out)  at (7,0) {$\ket {a_1 b_1+\cdots+a_n b_n}$};

	\node       	    at (4,0.3) {$\ldots$};
	\node	(OutT13) at (2.65,0) {};
	\node	 at (3.75,0) {} edge[-] (OutT13);
	\node	(InT23) at (5.35,0) {};
	\node	 at (4.25,0) {} edge[-] (InT23);
	\node	 at (6.15,0) {} edge[-] (out);
	
	\draw	(bn) -| (4.65,0.5) |- (5.25,0.5);
	\draw	(6.25,0.5) -| (6.75,1.5) |- (bnout);
	\draw[dashed] (bnin) -- (bn);
	\draw	(an) -| (4.8,1.5) |- (5.25,1);
	\draw	(6.25,1) -| (6.5,1.5) |- (anout);
	\draw[dashed] (anin) -- (an);
	
	\draw	(b1) -| (1.15,0.5) |- (1.75,0.5);
	\draw	(2.75,0.5) -| (3.25,1.5) |- (b1out);
	\draw[dashed] (b1out) -- (b1final);
	\draw	(a1) -| (1.3,1.5) |- (1.75,1);
	\draw	(2.75,1) -| (3,1.5) |- (a1out);
	\draw[dashed] (a1out) -- (a1final);

	\end{tikzpicture}
		\caption{Computing the inner product on a superposition of inputs. \label{fig:ip}}
\end{center}
\end{figure}

Figure~\ref{fig:ip} shows a reversible circuit for computing the inner product between two vectors. This is simply done by replacing the products by Toffoli gates. This quantum circuit is also itself reversible; applying the circuit twice leads to identity. In total, the inner product of two $n$-bit vectors can be computed, on a superposition of inputs, using $n$ Toffoli gates. Therefore, the computation of such a matrix vector multiplication requires at most $n^2$ Toffoli gates.

\paragraph{Matrix Multiplication.}
In the same fashion it is possible to compute matrix products in the quantum setting. Each column of the matrix can be computed using matrix vector multiplication, which, in turn, can be implemented using $n^2$ Toffoli gates. In total, a reversible quantum circuit for computing matrix multiplication on a superposition of inputs requires at most $n^3$ Toffoli gates, for square matrices.

\par \smallskip \noindent
Utilizing this quantum circuit for inner product between two vectors the quantum oracle for consistency checking can be constructed. Despite the fact that this naive quantum matrix multiplication is computed in $O(n^3)$, time greater than $O(n^{\omega})$ where $\omega \sim 2.376$ is the current classical complexity of matrix multiplication with the Coppersmith-Winograd algorithm, quantizing this computation will result in a lower quantum computational time for the classical \MQb{} problem. {\tt ClassicalBooleanSolve}, as well as {\tt SparseLinearSystemSolver} and the provided subroutines are analyzed in the black box model, where matrix multiplication such as $x \rightarrow Ax$ for a vector $x$ and a matrix $A$ are given by black boxes. We have provided the above construction to assure that such computations can be carried out by a quantum computer, reversibly and without entanglement concerns.

%% file: complexityanalysis.tex
We can now study the complexity of {\tt QuantumBooleanSolve} (Section \ref{sec:GroverBooleanSolve}). This analysis consists of constructing the quantum oracle $\mathcal{QBS}$ implementing $F^{{\rm cons}}_{F,k}: \{0,1\}^k \rightarrow \{0,1\}$, which on input $a \in \mathds{F}_2^k$ specializes the polynomial system $F$ and indicates the consistency of the associated Macaulay matrix $M_a$ of appropriate degree. This can be done by analyzing {\tt SparseLinearSystemSolver} and any associated subroutines separately, either illustrating the equivalence between the complexity of the classical function and the quantum circuit or proving that the quantum circuit is more efficient. \smallskip

For example, the quantization of the subroutine {\tt RandomSol} would consist of constructing a quantum circuit $\mathcal{QRS}$. {\tt RandomSol} takes as input a matrix $A \in \mathds{F}^{n \times n}$, a vector $b \in \mathds{F}^{n \times 1}$, a polynomial $f(z) \in \mathds{F}[z]$ with $f(0) \neq 0$ (and $\mathcal{L} \subset \mathds{F}$, which is in the case of \MQb{} a field extention of $\mathds{F}_2$. We would build 
\begin{eqnarray*}
\mathcal{QRS}: \ket{a_{11}} \ldots \ket{a_{nn}} \ket{f[0]^{-1}} \ket{f[1]} \ldots \ket{f[n]} \ket{r} \ket {w_1} \ldots \ket{w_n} \ket {b_1} \ldots \ket{b_n} \ket{0} \ldots \ket{0} \rightarrow \\
\ket{a_{11}} \ldots \ket{a_{nn}} \ket{f[0]^{-1}} \ket{f[1]} \ldots \ket{f[n]} \ket{r} \ket {w_1} \ldots \ket{w_n} \ket {b_1} \ldots \ket{b_n} \ket{0} \ldots \ket{0} \ket{b'_1} \ldots \ket{b'_r} \ket{x_1} \ldots \ket{x_n} \ket{s_A}
\end{eqnarray*} 
This quantum circuit takes as input the elements of the matrix $A$, the coefficients of the function $f$, the elements of the random vector $w$, $r=\text{deg}(f)$, and the elements of the vector $b$, as well as wires for computation space, and returns $b'_r=b+Aw$, $x=-\sum_{i=1}^n\frac{f[i]}{f[0]}A_r^ib'_r$, a boolean $s_A$ which takes the value of 1 if $Ax=b$ and 0 otherwise, along with the input for reversibility.

\begin{theorem} The quantum circuit $$\mathcal{QRS}: \ket{A} \ket{f} \ket{w} \ket{r} \ket{b} \ket{0} \ldots \ket{0} \rightarrow \ket{A} \ket{f}\ket{w}\ket{r}\ket{b} \ket{0} \ldots \ket{0} \ket{b'} \ket{x} \ket{s_A}$$ implementing {\tt RandomSol} requires $O(n^3+2n^2+3n+1)$ quantum gates to compute. In the black-box model, when provided with an oracle to compute matrix-vector and matrix-matrix products, $\mathcal{QRS}$ requires $O(r)$ evaluations of the black box, and $O(nr)$ operations in the base field $\mathds{F}$, which is equivalent to the classical complexity of {\tt RandomSol}.
\end{theorem}

A proof of the above theorem is fairly straightforward when directly analyzing a quantum analogue of the classical algorithm provided above for {\tt RandomSol}. It is clear that steps 4, 6, and 7 of {\tt RandomSol} are the only steps computed by the $\mathcal{QRS}$ quantum circuit. Firstly, step 4 consists of the computation of $b'_r=(b'_1,\ldots,b'_r)=b+Aw$, the first $r$ entries of the vector $b'$. This is merely matrix-vector multiplication and vector-vector addition; we compute the $r$ entries of $b'$ with $rn$ T gates for multiplication and $r$ CNOT gates for addition, totaling $O(rn+r) \leq O(n^2+n)$ quantum gates. In the black box model, we have 1 oracle query for matrix-vector multiplication and $O(r)$ field operations for addition of two vectors. Secondly, step 6 consists of computing for $i = 1 \ldots n$ the matrix-vector product $(A) \cdotp (A^{i-1}b')$ with $n^2$ T gates, followed by the computation of $\frac{f_i}{f_0}$ via one T-gate, and computing the $i$th term of the sum with an additional T gate. This is $O(n(n^2+2)+1)$ quantum gates to compute $x$ when we consider the additional NOT gate at the end of the computation. In the black box model, we have $O(n)$ black box matrix-vector product queries and $O(nr)$ field operations for the sum. Finally, the equality test conducted in step 7 consists of computing the matrix-vector product $Ax$ with $n^2$ T gates, followed by $n$ CNOT gates to compute, element by element, $(Ax)_i \oplus b_i$, and then one $T_{n+1}$ gate to compute the value $s_A$. In the black box model, this is 1 call to the matrix-vector product oracle. Therefore, we have established the equivalence of the classical complexity of the subroutine {\tt RandomSol} with the quantum oracle implementing the function $\mathcal{QRS}$ in the black-box model. \smallskip

Similar arguments demonstrate the equivalence of {\tt SparseLinearSystemSolver} as well as the entire quantum circuit $\mathcal{QBS}$. Due to the equivalence of the classical and quantum consistency checks in the black-box model, it is straightforward to adapt Theorem \ref{bs:cplx} to {\tt QuantumBooleanSolve}, as follows. 
\begin{theorem}\label{qbs:cplx}
Let  $\theta, 2 \leq \theta \leq 3$ is such that any two $n \times n$ matrices can be multiplied in $O(n^{\theta})$ operations in the underlying field. For any $\epsilon >0$, and $\alpha \geq 1$ and sufficiently large $m =\lceil \alpha n \rceil$, testing  the consistency of all Macaulay matrices in ${\tt QuantumBooleanSolve(m,n,k)}$ requires the:
\begin{itemize}
\item evaluation of $O(2^{(\frac{1 - \gamma}{2}+ \theta F_{\alpha}(\gamma) + \epsilon) n})$ quantum gates in the deterministic variant;
\item evaluation, on average, $O(2^{(\frac{1 - \gamma}{2} +2 F_{\alpha}(\gamma) + \epsilon) n})$ quantum gates in the probabilistic variant,
\end{itemize}
where $\gamma = 1 - \frac{k}{n}$, $F_{\alpha}(\gamma)= - \gamma \log_2(D^D(1-D)^{(1-D)})$ with $D=M(\frac{\alpha}{\gamma})$ and $M(x) = -x + \frac{1}{2} + \frac{1}{2} \sqrt{2x^2 - 10x - 1 + 2(x+2) \sqrt{x(x+2)}}$ and
\end{theorem}

The above complexity is obtained through the full evaluation of the cost of the consistency check oracle, $\mathcal{QBS}$, which is equivalent to the cost of the classical consistency check in the black box model. The quantum circuit for $\mathcal{QBS}$ can then be run in superposition over all generated Macaulay matrices, $\sum_{a \in \mathds{F}_2^k} \ket {{\mathbf M}_{{\bf a}}}$. Amplitude amplification is then utilized, as in Grover's algorithm, to determine the ${\bf a} \in \mathds{F}_2^k$ such that ${\mathbf M}_{{\bf a}}$ is inconsistent. \smallskip 

If we are guaranteed only one input ${\bf a} \in \mathds{F}_2^k$ is such that the generated Macaulay matrix ${\mathbf M}_{{\bf a}}$ is inconsistent, the algorithm requires $O(2^{k/2})=O(2^{(\frac{(1 - \gamma)n}{2}})$ evaluations of the quantum circuit $\mathcal{QBS}$ implementing $F^{{\rm cons}}_{F,k}$ for $F \in \polring{\F_2}{x}{n}^m$ as well as the diffusion gate $D$ for Grover's algorithm. When we have more than one ${\bf a} \in \mathds{F}_2$ such that ${\mathbf M}_{{\bf a}}$ is inconsistent, amplitude amplification must be run 
$$
O\left(\frac{\pi}{4} \sqrt{\frac{2^k}{\vert {\bf a} \in \mathds{F}^k_2 \text{ : } {\mathbf M}_{{\bf a}}\text{ inconsistent}\vert}}\right)
$$ times to recover such ${\bf a} \in \mathds{F}_2^k$.  \smallskip

As in the classical analysis of {\tt ClassicalBooleanSolve}, in the case that the Macaulay matrices are found to be inconsistent, the full system $F$ may be consistent. We therefore must determine the remainder of the solution, once we have found ${\bf a} \in \mathds{F}_2^k$ such that $M_{\bf a}$ is inconsistent. This exhaustive search can be performed using Grover's algorithm with a quantum oracle for the specialized system $\tilde{F}_a$, where if $a = (y_1,\ldots,y_k)$ we have $\tilde{F}=(\tilde{f}_1,\ldots,\tilde{f}_m)=(f_1(x_1,\ldots,x_{n-k},y_1,\ldots,y_k),\ldots,f_m(x_1,\ldots,x_{n-k},y_1,\ldots,y_k))$. Similarly to the classical analysis, we find that an overwhelming amount of computational cost is the consistency check performed by $\mathcal{QBS}$; the cost of the second exhaustive search, performed over the remaining $n-k$ variables, is negligible. By definition of strong semi-regularity, the number of such searches is bounded by $O(2^{(1 - 2 \gamma + 2F_{\alpha}(\gamma)_\epsilon)n})$, and therefore the cost of the second exhaustive search is bounded by the cost of the consistency check.\smallskip 

To derive the asymptotic complexity, we now minimize (for example, numerically) the exponents stated in Theorem \ref{qbs:cplx}. 
\begin{lemma} 
Let the notations be as in Theorem  \ref{qbs:cplx} and  $\alpha=1$. Then, the function 
$\frac{(1-\gamma)}{2}+\theta F_{\alpha}(\gamma)$ is bounded by: 
\begin{itemize}
\item $0.477=1-0.523$, when $\theta=3$ and $\gamma=0.1$,
\item $0.47=1-0.53$, when $\theta=2.376$ and $\gamma=0.13$,
\item $0.462=1-0.538$, when $\theta=2$ and $\gamma=0.17$
\end{itemize}    
\end{lemma} 
It can be remarked that the value of $\theta$ has a minimal impact on the bounds provided in the Lemma below; less than in the classical setting (see Section \ref{BooS}).  Note that these results can be extended to any $\alpha\geq 1$.
\smallskip 

To assure the reader that such computations can be performed on a quantum computer, we have provided a naive matrix-vector and matrix-matrix product circuit computed via inner product in the previous section. Finally, in summary, we have:

\begin{theorem} 
{\tt QuantumlBooleanSolve} is correct and solves \MQb{}.
If $m=n$, then -- for any $\epsilon >0$ --  the deterministic variant of the algorithm requires to evaluate 
$O(2^{(0.47+\epsilon)n})$ quantum gates provided 
that the system is $0.13$-strong semi-regular.  The Las-Vegas probabilistic variant  requires to evaluate an expected number 
of $O(2^{(0.462+\epsilon})n )$ quantum gates if the system is $0.17$-strong semi-regular.
\end{theorem}

This theorem follows directly from the equivalence of the classical and quantum complexity of the consistency checks, as well as the above Lemma 2.

This complexity should be directly compared to the ideal of a pure quadratic speed-up over the classical complexity of {\tt ClassicalBooleanSolve}, which is $O(2^{(0.396n)})$, as well as a quadratic speed-up on the classical approximation algorithm from \cite{NewMQ} which is $O(2^{(0.438n)})$. Note that none of these complexities have been obtained so  far and are thus  an open challenge.

%% file: FastQuantumAlgo.bbl
\begin{thebibliography}{10}
\providecommand{\url}[1]{\texttt{#1}}
\providecommand{\urlprefix}{URL }

\bibitem{DBLP:journals/iacr/AlbrechtCFFP14}
Albrecht, M.R., Cid, C., Faug{\`{e}}re, J., Fitzpatrick, R., Perret, L.:
  Algebraic algorithms for {LWE} problems. {IACR} Cryptology ePrint Archive
  2014,  1018 (2014), \url{http://eprint.iacr.org/2014/1018}

\bibitem{DBLP:conf/icalp/AroraG11}
Arora, S., Ge, R.: New algorithms for learning in presence of errors. In:
  Aceto, L., Henzinger, M., Sgall, J. (eds.) Automata, Languages and
  Programming - 38th International Colloquium, {ICALP} 2011, Zurich,
  Switzerland, July 4-8, 2011, Proceedings, Part {I}. Lecture Notes in Computer
  Science, vol. 6755, pp. 403--415. Springer (2011),
  \url{https://doi.org/10.1007/978-3-642-22006-7_34}

\bibitem{bardet2013complexity}
Bardet, M., Faug{\`e}re, J.C., Salvy, B., Spaenlehauer, P.J.: On the complexity
  of solving quadratic boolean systems. Journal of Complexity  29(1),  53--75
  (2013)

\bibitem{DBLP:journals/siamcomp/BennettBBV97}
Bennett, C.H., Bernstein, E., Brassard, G., Vazirani, U.V.: Strengths and
  weaknesses of quantum computing. {SIAM} J. Comput.  26(5),  1510--1523
  (1997), \url{http://dx.doi.org/10.1137/S0097539796300933}

\bibitem{DBLP:journals/jsc/BerbainGP09}
Berbain, C., Gilbert, H., Patarin, J.: {QUAD:} {A} multivariate stream cipher
  with provable security. J. Symb. Comput.  44(12),  1703--1723 (2009),
  \url{http://dx.doi.org/10.1016/j.jsc.2008.10.004}

\bibitem{opac-b1128738}
Bernstein, D.J., Buchmann, J., Dahmen, E. (eds.): Post-quantum cryptography.
  Mathematics and Statistics Springer-11649; ZDB-2-SMA, Springer Berlin
  Heidelberg, Berlin, Heidelberg (2009),
  \url{http://opac.inria.fr/record=b1128738}

\bibitem{BFP09b}
Bettale, L., Faug\`ere, J.C., Perret, L.: {Hybrid Approach for Solving
  Multivariate Systems over Finite Fields}. Journal of Mathematical Cryptology
  3(3),  177--197 (2010), \url{http://www-salsa.lip6.fr/~jcf/Papers/JMC2.pdf}

\bibitem{FBP12}
Bettale, L., Faug{\`e}re, J.C., Perret, L.: {Solving Polynomial Systems over
  Finite Fields: Improved Analysis of the Hybrid Approach}. In: Proceedings of
  the 37th International Symposium on Symbolic and Algebraic Computation. pp.
  67--74. ISSAC '12, ACM, New York, NY, USA (2012),
  \url{http://www-polsys.lip6.fr/~jcf/Papers/FBP12.pdf}

\bibitem{PKC:BetFauPer11}
Bettale, L., Faug{\`e}re, J.C., Perret, L.: Cryptanalysis of multivariate and
  odd-characteristic {HFE} variants. pp. 441--458

\bibitem{CHES:BCCCNS10}
Bouillaguet, C., Chen, H.C., Cheng, C.M., Chou, T., Niederhagen, R., Shamir,
  A., Yang, B.Y.: Fast exhaustive search for polynomial systems in $f_2$. pp.
  203--218

\bibitem{MR1947332}
Brassard, G., H{\o}yer, P., Mosca, M., Tapp, A.: Quantum amplitude
  amplification and estimation. In: Quantum computation and information
  ({W}ashington, {DC}, 2000), Contemp. Math., vol. 305, pp. 53--74. Amer. Math.
  Soc., Providence, RI (2002), \url{http://dx.doi.org/10.1090/conm/305/05215}

\bibitem{DBLP:journals/jsc/Buchberger06a}
Buchberger, B.: Bruno {B}uchberger's {PhD} thesis 1965: An algorithm for
  finding the basis elements of the residue class ring of a zero dimensional
  polynomial ideal. Journal of Symbolic Computation  41(3-4),  475--511 (2006)

\bibitem{BCLA82}
Buchberger, B., Collins, G.E., Loos, R.G.K., Albrecht, R.: Computer algebra
  symbolic and algebraic computation. SIGSAM Bull.  16(4),  5--5 (1982)

\bibitem{DBLP:conf/space/2016}
Carlet, C., Hasan, M.A., Saraswat, V. (eds.): Security, Privacy, and Applied
  Cryptography Engineering - 6th International Conference, {SPACE} 2016,
  Hyderabad, India, December 14-18, 2016, Proceedings, Lecture Notes in
  Computer Science, vol. 10076. Springer (2016),
  \url{https://doi.org/10.1007/978-3-319-49445-6}

\bibitem{NISTPQ}
Chen, L., Jordan, S., Liu, Y.K., Moody, D., Peralta, R., Perlner, R.,
  Smith-Tone, D.: Report on post-quantum cryptography. Reasearch report NISTIR
  8105, NIST (2003),
  \url{http://csrc.nist.gov/publications/drafts/nistir-8105/nistir_8105_draft.pdf}

\bibitem{DBLP:conf/asiacrypt/ChenHRSS16}
Chen, M., H{\"{u}}lsing, A., Rijneveld, J., Samardjiska, S., Schwabe, P.: From
  5-pass \emph{MQ} -based identification to \emph{MQ} -based signatures. In:
  Cheon, J.H., Takagi, T. (eds.) Advances in Cryptology - {ASIACRYPT} 2016 -
  22nd International Conference on the Theory and Application of Cryptology and
  Information Security, Hanoi, Vietnam, December 4-8, 2016, Proceedings, Part
  {II}. Lecture Notes in Computer Science, vol. 10032, pp. 135--165 (2016),
  \url{https://doi.org/10.1007/978-3-662-53890-6_5}

\bibitem{cryptoeprint:2017:680}
Chen, M.S., Hülsing, A., Rijneveld, J., Samardjiska, S., Schwabe, P.: Sofia:
  Mq-based signatures in the qrom. Cryptology ePrint Archive, Report 2017/680
  (2017), \url{http://eprint.iacr.org/2017/680}

\bibitem{FOPT10}
Faug\`ere, J.C., Otmani, A., Perret, L., Tillich, J.P.: {Algebraic
  Cryptanalysis of {McEliece} Variants with Compact Keys}. In: Proceedings of
  Eurocrypt 2010. Lecture Notes in Computer Science, vol. 6110, pp. 279--298.
  Springer Verlag (2010),
  \url{http://www-salsa.lip6.fr/~jcf/Papers/Eurocrypt2010.pdf}

\bibitem{faugere:hal-01064687}
Faug{\`e}re, J.C., Perret, L., De~Portzamparc, F.: {Algebraic Attack against
  Variants of {M}cEliece with {G}oppa Polynomial of a Special Form}. In:
  {Advances in Cryptology Asiacrypt 2014}. Kaohsiung, Taïwan (Sep 2014),
  \url{http://hal.inria.fr/hal-01064687}

\bibitem{C:FauJou03}
Faug{\`e}re, J.C., Joux, A.: Algebraic cryptanalysis of hidden field equation
  ({HFE}) cryptosystems using gr{\"o}bner bases. pp. 44--60

\bibitem{DBLP:conf/issac/Gall14a}
Gall, F.L.: Powers of tensors and fast matrix multiplication. In: Nabeshima,
  K., Nagasaka, K., Winkler, F., Sz{\'{a}}nt{\'{o}}, {\'{A}}. (eds.)
  International Symposium on Symbolic and Algebraic Computation, {ISSAC} '14,
  Kobe, Japan, July 23-25, 2014. pp. 296--303. {ACM} (2014),
  \url{http://doi.acm.org/10.1145/2608628.2608664}

\bibitem{GJ79}
Garey, M.R., Johnson, D.S.: Computers and Intractability: A Guide to the Theory
  of {NP}-Completeness. W. H. Freeman (1979)

\bibitem{DBLP:books/daglib/0031325}
von~zur Gathen, J., Gerhard, J.: Modern Computer Algebra {(3. ed)}. Cambridge
  University Press (2013)

\bibitem{bananas}
Giesbrecht, M., Lobo, A., Saunders, B.D.: Certifying inconsistency of sparse
  linear systems. In: Proceedings of the 1998 international symposium on
  Symbolic and algebraic computation. pp. 113--119. ACM (1998)

\bibitem{grover1996fast}
Grover, L.K.: A fast quantum mechanical algorithm for database search. In:
  Proceedings of the twenty-eighth annual ACM symposium on Theory of computing.
  pp. 212--219. ACM (1996)

\bibitem{DBLP:journals/tcs/IvanyosS17}
Ivanyos, G., Santha, M.: Solving systems of diagonal polynomial equations over
  finite fields. Theor. Comput. Sci.  657,  73--85 (2017),
  \url{https://doi.org/10.1016/j.tcs.2016.04.045}

\bibitem{EC:KipPatGou99}
Kipnis, A., Patarin, J., Goubin, L.: Unbalanced oil and vinegar signature
  schemes. pp. 206--222

\bibitem{NewMQ}
Lokshtanov, D., Paturi, R., Tamaki, S., Williams, R., Yu, H.: Beating brute
  force for systems of polynomial equations over finite fields, to appear, 27th
  ACM-SIAM Symposium on Discrete Algorithms (SODA 2017)

\bibitem{NISTsubmit}
{NIST}: Proposed submission requirements and evaluation criteria for the
  post-quantum cryptography standardization process ({DRAFT}).,
  \url{http://csrc.nist.gov/groups/ST/post-quantum-crypto/documents/call-for-proposals-draft-aug-2016.pdf}

\bibitem{DBLP:books/hal/Perret16}
Perret, L.: Bases de Gr{\"{o}}bner en Cryptographie Post-Quantique.
  (Gr{\"{o}}bner bases techniques in Quantum-Safe Cryptography) (2016),
  \url{https://tel.archives-ouvertes.fr/tel-01417808}

\bibitem{westerbaan2016solving}
Westerbaan, B., Schwabe, P.: Solving binary mq with grover's algorithm. In:
  Carlet, C.; Hasan, A.; Saraswat, V.(ed.), Security, Privacy, and Advanced
  Cryptography Engineering: 6th International Conference, SPACE 2016,
  Hyderabad, India, December 14-18, 2016. pp. 303--322. Berlin: Springer-Verlag
  (2016)

\bibitem{wiedemann1986solving}
Wiedemann, D.: Solving sparse linear equations over finite fields. IEEE
  transactions on information theory  32(1),  54--62 (1986)

\end{thebibliography}
